\documentclass[epj]{svjour}

\usepackage{graphicx}
\usepackage{fancyhdr}
\def\makeheadbox{{
 }}
\def\mytitle{My title} 
\def\myauthors{My name}  
\def\mytype{My type of session}
\def\mysession{My session}

\def\mytitle{$CP$-violation in $b \to s$ Penguin Decays at BaBar} 
\def\myauthors{Nitesh Soni}    
\def\mytype{Contributed Talk}    
\def\mysession{Flavor Physics}

\begin{document}
\title{CP-violation in $b \to s$ Penguin Decays at BaBar}
\author{Nitesh Soni
\thanks{\emph{Email:} nitesh@slac.stanford.edu}, %
 for the {\it BABAR} Collaboration
}                     
%
%
\institute{School of Physics and Astronomy, University of Birmingham, Edgbaston, Birmingham - B15 2TT (U.K.)
}
%
\date{October, 2007}
\abstract{
We present the new and updated $BABAR$ measurements of $CP$-violation studies for many $b\to s $ penguin decay modes. We report the first observation of mixing-induced $CP$-violation in $B^0 \to \eta ^{'} K^0$ with a significance (including systematic uncertainties) of 5.5$\sigma$. We also present the first observation of the decay $B^0 \to \rho ^0 K^0$. Using the time-dependent Dalitz plot analysis of $B^0\to K^+K^- K^0$ decay, the $CP$-parameters $\mathcal A_{CP}$ and $\beta _{eff}$ are measured with $4.8\sigma$ significance, and we reject the solution near $\pi /2 - \beta _{eff}$ at $4.5\sigma$. We also present the updated measurements of $CP$-violating parameters for $B^0 \to K_S^0 \pi ^0$, $K_S^0K_S^0K_S^0$ and $\pi ^0 \pi ^0 K_S^0$ decays. An updated measurements of the $CP$-violating charge asymmetries for $B^{\pm}\to \eta ^{'} K^{\pm}$, $\eta K^{\pm}$, $\omega K^{\pm}$ decays are also presented. The measurements are based on the data sample recorded at the $\Upsilon (4S)$ resonance  with $BABAR$ detector at the PEP-II $B$-meson Factory at SLAC.
} 
\maketitle
\section{Introduction}
\label{intro}
Measurements of time-dependent $CP$-asymmetries in $B^0$ meson decays through Cabibbo-Kobayashi-Maskawa (CKM) favoured $b\to c \bar {c}s$ amplitudes \cite{IntroRef1} have provided crucial tests of the mechanism of $CP$-violation in the standard model (SM) \cite{IntroRef2}. A major goal of $B$ Factory experiments is now to search for indirect evidence for New Physics (NP). One strategy is to compare the measured value of the $CP$-violation (CPV) parameters from $b \to c\bar cs$ to independent determinations of the same quantities, from processes that are sensitive to the contributions of NP effects through loop (penguin) diagrams. In the decay chain $\Upsilon (4S) \to B^0 \overline B^0$, one of the two $B$ mesons decays into a $CP$ eigenstate $f_{CP}$ at time $t_{CP}$, and other decays into a flavor specific state $f_{\rm tag}$ at time $t_{\rm tag}$. The CPV parameters are measured using time-dependent decay rate
\begin{eqnarray}
\nonumber  \lefteqn{\mathcal P (\Delta t) = \frac {e^{-|\Delta t|/\tau}}{4\tau}} \\
   && \times [ 1+q\{ S_f\sin (\Delta m_d\Delta t) - C_f \cos(\Delta m_d\Delta t)\} ]
\end{eqnarray}
where $\Delta t = t_{CP} - t_{\rm tag}$, $\tau$ is the $B$ lifetime, $\Delta m_d$ is the $B^0\overline B^0$ mixing frequency and $q = +1(-1)$ when $f_{\rm tag} = B^0 (\overline B^0)$. The parameters $S_f$ and $C_f$ represent the mixing-induced and direct CPV respectively.
For penguin-dominated processes $b\to s q\bar q$ $(q = u,d,s)$, $S_f$ and $C_f$ are expected to be consistent with the values from $b\to c\bar c s$ decays( namely, $C_f \sim 0$ and $S_f \sim \sin 2\beta$). Additional CKM suppressed contributions to the amplitude can induce only small deviations from the expectation. On the other hand, additional loop contributions from NP processes may produce large observable deviations~\cite{penguin}. In this paper, we present new and updated measurements of $b\to s$ penguin decays. 
\section{Analysis Overview}
\label{sec:1} 
The updated and new measurements presented here are based on the data collected with the $BABAR$ detector~\cite{babar} at the PEP-II asymmetric energy $e^+e^-$ collider, located at SLAC, USA. Unless explicitly stated, all the analyses are presented with a data sample corresponds to $383 \pm 4$ million $B\overline B$ pairs recorded at $\Upsilon (4S)$ resonance (Center-of-mass energy (CM) $\sqrt{s}~=~10.58$ GeV) and $37~fb^{-1}$ data collected about $40~{\rm MeV}$ below the $\Upsilon (4S)$ resonance (``off-resonance''). The $BABAR$ detector provides the charged particle tracking through a combination of a 5-layer silicon vertex tracker (SVT) and a 40-layer drift chamber (DCH), both operating within a 1.5 T magnetic field generated by a superconducting solenoidal magnet. Photons are identified in an electromagnetic calorimeter (EMC) surrounding a detector of internally reflected Cherenkov light (DIRC), which associates Cherenkov photons with tracks for particle identification (PID). Muon candidates are identified with the use of the instrumented flux return (IFR) of the solenoid.

To discriminate between signal and background, two independent kinematic variables, the beam energy substituted mass $m_{\rm ES} \equiv \sqrt {s/4 - ({p_B^*})^2}$ and energy difference $\Delta E \equiv E_B^* - \sqrt {s}/2$ are used. Here $E_B^*$ and $p_B^*$ are the CM energy and 3-momentum of $B$ candidate, respectively. For signal decays, $m_{\rm ES}$ ($\Delta E$) peak at nominal $B$-mass (at zero) with a resolution of a few (a few tens of) MeV. We use event topology to reject the continuum background (arises primarily from random combinations of particles in $e^+e^- \to q\bar q$ events ($q = u, d, s, c$) events which have a ``jet-like'' structure against the real $B$ events which are more uniform. To improve the signal-background separation, some of these variables are combined in a multivariate algorithm (a Fisher discriminant or Neural Network etc.) which is trained on the signal monte carlo and $q\bar q$ background and used in the fits.

The physical quantities (e.g. signal yields or $CP$-parameters) are determined from extended and unbinned maximum likelihood (ML) fits. The likelihood function include the probability density functions (PDF) whose shapes are based on MC or data (sidebands or off-resonance). Their most important parameters are left free. In addition to signal and continuum, $B$-background categories are also included in the fits. Then we calculate the systematic errors arising from several sources e.g. variation of signal PDF shape parameters within their errors, $B$-background estimation, the effect of interference etc. In all the reported measurements of $CP$-violation below, the statistical errors dominates.
\section{Results}
\label{sec:2}
\subsection{$B^0 \to \eta ^{'} K^0$}
\label{subsec:1} 
Theoretically, it has the large Branching Fractions ($65 \times 10^{-6}$) as compared to other penguin decay modes. Here we update our previous measurements. In addition to the $B^0 \to \eta ^{'} K_S^0$ decays used previously, we now also include the decay $B^0 \to \eta ^{'} K_L^0$. The $B$-daughter candidates are reconstructed through their decays $\pi ^0 \to \gamma \gamma$, $\eta \to \gamma \gamma (\eta _{\gamma \gamma})$, $\eta \to \pi ^+ \pi ^- \pi ^0 (\eta _{3 \pi})$, $\eta ^{'} \to \eta _{\gamma \gamma} \pi ^+ \pi ^-$, $\eta ^{'} \to \eta _{3 \pi} \pi ^+ \pi ^-$, $\eta ^{'} \to \rho ^0 \gamma$. Only $\eta ^{'} \to \eta _{\gamma \gamma} \pi ^+ \pi ^-$ mode is used for the $\eta ^{'} K_L^0$. The detail of this analysis is described in Ref.~\cite{etaks}. We obtain the CPV parameters and signal yields for each channel from a ML fit with the input observables $\Delta E, ~m_{\rm ES}$, $Fisher$ and $\Delta t$. Finally we obtain the $CP$-violations parameters, $S = 0.58 \pm 0.10 \pm 0.03$ and $C = -0.16 \pm 0.07 \pm 0.03$. The data used is 2.1 times as large as that of previous measurements~\cite{etaks-old}. We observe mixing-induced $CP$-violation in $B^0 \to \eta ^{'} K^0$ decay with a significance of 5.5 standard deviations (systematic uncertainties included). The results for direct $CP$-violation is 2.1 standard deviations from zero.
\subsection{$B^0 \to \rho ^0 K^0$}
\label{subsec:2}
This analysis is reported with the $227 \times 10^6 B\overline B$ pairs. We take a quasi-two body approach, restricting ourselves to the region of the $B^0 \to \pi ^+ \pi ^- K_S^0$ Dalitz plot (DP) dominated by the $\rho ^0$ and treating other $B^0 \to \pi ^+ \pi ^- K_S^0$ contributions as a non-interfering background. The effects of interference with other resonances are estimated and taken as a systematic uncertainties. The $B^0$ candidates are reconstructed from the combinations of $\rho ^0 \to \pi ^+ \pi ^-$ and $K_S^0 \to \pi ^+ \pi ^-$. An unbinned extended ML fit with the input of $m_{\rm ES}, \Delta E, NN,$ $\cos \theta _{\pi ^+}$, $m_{\pi\pi}$, $\Delta t$ is used to extract the $CP$-asymmetry and branching fractions; here $NN$ is the neural network to enhance discrimination between signal and continuum background, $\cos \theta _{\pi ^+}$ is the angle between the $K_S^0$ and the $\pi ^+$ from the $\rho ^0$ in the $\rho ^0$ meson's CM frame and $m_{\pi\pi}$ is the mass of pion pairs. The fit yields $111 \pm 19$ signal events, we calculate the Branching Fractions $(B^0 \to \rho ^0 K^0) = (4.9 \pm 0.8 \pm 0.9) \times 10^{-6}$, where the first error is statistical and the second systematic. The likelihood ratio between the fit result of 111 signal events and the null hypothesis of zero signal shows that this excluded at the $8.0\sigma$ level. When additive systematic effects are included, we exclude the null hypothesis at the $5.0\sigma$ level. This is the first observation of the $B^0 \to \rho ^0 K^0$. The fit for the $CP$-parameters gives $S = 0.20 \pm 0.52 \pm 0.24$ and $C = 0.64 \pm 0.41 \pm 0.20$. The detail of this analysis can be seen at Ref.~\cite{rhoks}.
\subsection{$B^0 \to K^+ K^- K^0$}
\label{subsec:3}
\begin{figure}
\includegraphics[width=0.50\textwidth,height=0.25\textwidth,angle=0]{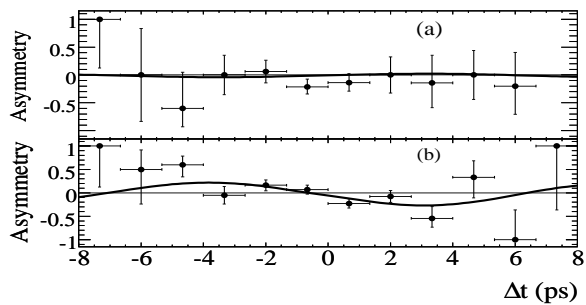}
\caption{The raw asymmetry between $B^0$- and $\overline B^0$-tagged events in (a) low-mass region and (b) the high-mass region. The curves are projections of corresponding fit results.}
\label{figure:2}       
\end{figure}

Previous measurements of $CP$-asymmetries in $B^0 \to K^+ K^- K^0$ have been performed separately for the events $K^+ K^-$ in the $\phi$ mass region, and for events excluding the $\phi$ region, neglecting interference effects among intermediate states ~\cite{kkk-old}. But now we have performed a time-dependent Dalitz plot analysis of $B^0 \to K^+ K^- K^0$ decay from which we extract the values of $CP$-violation parameters by taking into account the complex amplitudes describing the entire $B^0$ and $\overline B^0$ Dalitz plots. We define the complex amplitude $\mathcal A$ for $B^0$ decay as a sum of isobar amplitudes ~\cite{amplitudes},
\begin{eqnarray}
\nonumber  \lefteqn{\mathcal{A}(m_{K^+K^-}, \cos \theta _{H}) = \sum_r{\mathcal A}_r}\\
   && =\sum_r c_r(1 + b_r)e^{i(\phi _r + \delta _r)}.f_r(m_{K^+K^-}, \cos\theta _{H}),
\end{eqnarray}
where the parameters $c_r$ and $\phi _r$ are the magnitude and phase of the amplitude of the component $r$, and we allow for different isobar coefficients for $B^0$ decays through the asymmetry parameters $b_r$ and $\delta _r$. Similarly, we can write the amplitude $\overline \mathcal{A}$ for $\overline {B^0}$ by replacing the plus signs with minus signs in above equation. Our isobar model includes resonant amplitudes $\phi ,~ f_0,~\chi _{c0}(1P),$ and $X_0(1550)$~\cite{isobarmodel}; non-resonant terms; and incoherent terms of $B^0$ decay to $D^- K^+$ and $D_s^- K^+$. For each resonant term, the function $f_r = F_r \times T_r \times Z_r$ describes the dynamical properties, where $F_r$ is the Blatt-Weisskkopf centrifugal barrier factor for the resonance decay vertex~\cite{fr}, $T_r$ is the resonant mass line shape, and $Z_r$ describes the angular distribution in the decay~\cite{zr}. The $CP$-asymmetries parameters $A_{CP,r}$(Rate Asymmetry) and $\beta _{eff,r}$ (Phase asymmetry) can be computed as 
\begin{eqnarray}
A_{CP,r} = \frac {|\bar{\mathcal A_r}|^2 -|{\mathcal A_r}|^2}{|\bar{\mathcal A_r}|^2 + |{\mathcal A_r}|^2} = \frac {-2b_r}{1+b_r^2},\\
\beta _{eff, r} =  \beta + \delta _r
\end{eqnarray}
We first extract the values of the parameters of the amplitude model, and measure the average $CP$-asymmetry in $B^0 \to K^+ K^- K^0$ decay over the entire Dalitz plot.  The fraction of events (${\mathcal F}_r$) obtained from the fit to whole Dalitz plot are shown in Table ~\ref{table:1}, and the $CP$-asymmetries results are summarised in Table~\ref{table:2}. The sum of the fractions is greater than one due to the interference. Using this model, we then measure the $CP$-asymmetries for $\phi K^0$ and $f_0 K^0$ decay channels, from a ``low mass''  analysis of events with $M_{K^+K^-} < 1.1$~${\rm GeV/c^2}$. Finally, we perform a ``high mass'' analysis to determine the average $CP$-asymmetry for events with $M_{K^+K^-} > 1.1$~${\rm GeV/c^2}$.  The raw asymmetry between $B^0$ - and $\overline B^0$-tagged events in (a) the low-mass region and (b) high-mass region are shown in Figure~\ref{figure:2}, and the obtained value of $CP$-parameters are given in Table~\ref{table:2}. The values of $\beta _{eff}$ and $A_{CP}$ obtained from the whole DP fit are consistent with SM expectations of $\beta \simeq 0.370$ rad, $ A_{CP} \simeq 0$. The significance of $CP$-violation is $4.8 \sigma$, and we reject the solution near $\pi /2 - \beta$ at $4.5 \sigma$. In the fit to the low-mass region of DP, we find $\beta _{eff}$ lower than the SM expectation about by $2\sigma$ while in the high-mass region $CP$-parameters are compatible with SM expectations and we observe $CP$-violation at the level of $5.1 \sigma$. The more detail of this analysis can be found in Ref.~\cite{kkk}.\\
\begin{table}
\begin{center}
\caption{The fit fractions ${\mathcal F}_r$ from the fit to the full $B^0 \to K^+ K^- K^0$ Dalitz Plot. Errors are statistical only.}
\label{table:1}       
\begin{tabular}{lc}
\hline\hline\noalign{\smallskip}
Isobar Model & ${\mathcal F}_r (\%)$   \\
\noalign{\smallskip}\hline\noalign{\smallskip}
$\phi K^0$ &$12.5\pm 1.3$ \\
$f_0 K^0$ & $40.2\pm 9.6$ \\
$X_0 (1550) K^0$ &  $4.1\pm 1.3$ \\
$(K^+ K^0 K^-)_{NR}$ & $112.0 \pm 14.9$ \\
$\chi _{c0} (1P) K^0$ & $3.0\pm 1.2$ \\
\noalign{\smallskip}\hline\hline
\end{tabular}
\end{center}\end{table}
\begin{table}
\begin{center}
\caption{The $CP$-asymmetries for $B^0 \to K^+K^-K^0$ for the entire DP, in the high mass region, and for $\phi K^0$ and $f_0 K^0$ in the low mass region. The first errors are statistical and second errors are systematics. }
\label{table:2}       
\resizebox{8cm}{!} {
\begin{tabular}{lcc}
\hline\hline\noalign{\smallskip}
 & $A_{CP}$ & $\beta _{eff}$(rad.)   \\
\noalign{\smallskip}\hline\noalign{\smallskip}
Whole DP & $-0.015 \pm 0.077 \pm 0.053$ & $0.352 \pm 0.076 \pm 0.026$\\\hline
High-mass & $-0.054 \pm 0.102 \pm 0.060$ & $0.436 \pm 0.087^{+0.055}_{-0.031}$ \\\hline
$\phi K^0$ & $-0.08 \pm 0.18 \pm 0.04$ & $0.11 \pm 0.14 \pm 0.06$ \\
$f_0 K^0$& $~~0.41 \pm 0.23 \pm 0.07$ & $0.14 \pm 0.15 \pm 0.05$ \\
\noalign{\smallskip}\hline\hline
\end{tabular}}
\end{center}\end{table}
\subsection{$B^0\to K_S^0 \pi^0 $}
\label{subsec:4}
A precise measurement of direct $CP$-asymmetry and branching fraction for $B^0 \to K^0 \pi ^0$ is important as it is possible to combine direct $CP$-asymmetries and branching fraction measurements for all the four $B \to K\pi$ modes to test precise sum rules. We have presented the updated measurement for this analysis. Since no charged particles are produced at the decay vertex of $B$ so we determine the decay point by constraining the $B$ production vertex to the interaction point (IP) in the plane orthogonal to the beam axis using only the $K_S^0 \to \pi ^+ \pi ^-$ trajectories. We account for possible biases due to the vertexing technique by comparing fits to a large simulated sample of IP-constrained (neglecting the $J/\psi$ contribution to the vertex and using $K_S^0$ trajectory only) and nominal $B^0\to J/\psi K_S^0$ events. The detailed analysis technique is discussed in Ref. ~\cite{kspizero}. We have measured the CPV parameters $C_{K_S^0 \pi ^0}~=~0.24\pm 0.15\pm 0.03$ and $S_{K_S^0 \pi ^0}~=~0.40\pm 0.23\pm 0.03$, and branching fraction $\mathcal {B}(B^0 \to K^0\pi ^0)~=~(10.34\pm 0.66 \pm 0.58)\times 10^{-6}$. The first errors are statistical and the second are systematic. These values are consistent with SM predictions and the experimental value of $\sin 2\beta$; and the results supersede $BABAR$ previous measurements.
\subsection{$B^0\to K_S^0 K_S^0 K_S^0$}
\label{subsec:5}
The decay {$B^0\to K_S^0 K_S^0 K_S^0$} is pure $CP$-even eigenstate and the theoretical uncertainty in the SM prediction of $\sin 2\beta _{eff}$ is less than 4\%. The Ref.~\cite{threeks} gives the whole detail of this analysis. The whole data sample is divided into two subsamples, one where all three $K_S^0$ mesons decays into the $\pi ^+ \pi ^-$ channel ($B_{CP(+-)}$) and another where one of the $K_S^0$ mesons decays into the $\pi ^0 \pi ^0$ channel, while the other two decay into the $\pi ^+ \pi ^-$ channel ($B_{CP(00)}$) . We form $\pi ^0 \to \gamma \gamma$ candidates from pairs of photon candidates in the EMC. For each $B_{CP(+-)}$ candidate $m_{\rm ES}$ and $\Delta E$ are the two independent kinematical variable used while for $B_{CP(00)}$ reconstructed $B^0$ mass $m_B$ and the missing mass $m_{\rm miss} = \sqrt{(q_{e^+e^-} - \tilde{q_B})^2}$ is used. Here $\tilde {q_B}$ is the four-momentum of the $B_{CP(00)}$ candidate after a mass constraint on the $B^0$ meson has been applied. Since no charged particles are produced at the decay vertex of $B$, we have applied the same technique to find the vertex as used in $B^0\to K_S^0 \pi ^0$ (explained in Section~\ref{subsec:4}). The measured value of time-dependant $CP$-asymmetries, $S = -0.71\pm 0.24 \pm 0.04$ and $C = 0.02\pm 0.21 \pm 0.05$, where the first errors are statistical and the second systematic. The statistical correlation between $S$ and $C$ is -14.1\%.  The statistical significance of $CP$-violation is $2.9\sigma$. These results agree well with the SM expectation and supersede our previous measurements.
\subsection{$B^0\to \pi^0 \pi^0 K_S^0$}
\label{subsec:6}
The {$B^0\to \pi^0 \pi^0 K_S^0$} final state is $CP$-even eigen state, regardless of any resonant substructure.  Just like {$B^0\to K_S^0 \pi ^0$} and {$B^0\to K_S^0 K_S^0 K_S^0$} analyses, the same technique is used to find the vertex of $B$ candidates. We measure the $CP$-violating asymmetries in this decay from a sample of approximately 227 million $B\overline B$ pairs. From an unbinned extended ML fit we obtain $S = 0.72\pm 0.71\pm 0.08$ and $C = 0.23\pm 0.52\pm 0.13$. When we fix the values of $-S$ to the average $\sin2\beta$ measured in $b\to c\bar c s$ modes, and $C$ to zero, and re-fit the data sample the negative likelihood changes by $2.2\sigma$. The detail of this analysis is described in Ref.~\cite{pipiks}.
\subsection{$B^{\pm}\to \eta ^{'} K^{\pm}, \eta K^{\pm}, \omega K^{\pm}$}
\label{subsec:7}
The decays $B \to \eta K$ are especially interesting since they are suppressed relative to the abundant $B\to \eta ^{'} K$ decays due to destructive interference between two penguin amplitudes. We present the updated measurement of $CP$-violating charge asymmetries ($\mathcal A _{ch}$) for $B^\pm$. For $B^\pm \to f^\pm$ decays, the direct $CP$-violation would correspond to a non-zero charge asymmetry defined as
\begin{eqnarray}
\mathcal A _{ch} \equiv \frac {\Gamma (B^- \to f^-) - \Gamma (B^+ \to f^+)}{\Gamma (B^- \to f^-) + \Gamma (B^+ \to f^+)}
\end{eqnarray}
Two decay chains are reconstructed for the $\eta (\to \gamma \gamma$ {\rm and}~$ \to \pi ^+\pi ^- \pi ^0$) and $\eta ^{'} (\to \pi ^+ \pi ^- \pi ^0 $ {\rm and}~$\rho ^0 (\to \pi ^+ \pi ^-) \gamma$) candidates. Then ML fit are done separately for each decay chain; the likelihood functions (including systematic errors) are then combined to provide the charge asymmetries. We obtain the $\mathcal A _{ch} (B^+ \to \eta K^+) = -0.22 \pm 0.11\pm 0.01$, $\mathcal A _{ch} (B^+ \to \eta ^{'}K^+) = 0.010 \pm 0.022\pm 0.006$ and $\mathcal A _{ch} (B^+ \to \omega K^+) = -0.01 \pm 0.07\pm 0.01$. We find no clear evidence for direct $CP$-violation charge asymmetries in these decays. The detail of this analysis can be found in Ref.~\cite{chargedb}.
\section{Conclusion}
\label{sec:4}
We presented new and updated $CP$-violation studies for many $b\to s $ penguin decay modes. We reported the first observation of mixing-induced $CP$-violation in $B^0 \to \eta ^{'} K^0$ with a significance (including systematic uncertainties) of 5.5$\sigma$. We also presented the first observation of the decay $B^0 \to \rho ^0 K^0$ and a measurement of $CP$-violating parameters. Using time-dependent Dalitz plot analysis of $B^0\to K^+K^- K^0$ decay, the $\mathcal A_{CP}$ and $\beta _{eff}$ are measured. The significance of $CP$-violation is $4.8\sigma$, and we reject the solution near $\pi /2 - \beta _{eff}$ at $4.5\sigma$. We have also presented the updated measurements of $CP$-violating parameters for $B^0 \to K_S^0 \pi ^0$, $K_S^0K_S^0K_S^0$ and $\pi ^0 \pi ^0 K_S^0$ decays. All results supersede previous $BABAR$ results. Finally, an updated measurements of the $CP$-violating charge asymmetries for $B^{\pm}\to \eta ^{'} K^{\pm}, \eta K^{\pm}, \omega K^{\pm}$ decays are presented. The individual results of $S \sim \sin 2\beta _{eff}$ are consistent with world average of $\sin 2\beta$ from $b\to c\bar c s$ but naive average of $\sin 2\beta _{eff}$ from all penguin modes is lower than $\sin 2\beta$~\cite{hfag} which is opposite to theoretical prediction. Finally, we do not find any evidence of direct $CP$-violation in the decay modes reported.

\end{document}